\documentclass[12pt]{article}
\usepackage{amssymb}
\usepackage{amsmath}
\usepackage{amstext}
\usepackage{graphicx,epsfig}
\usepackage{epsfig}
\usepackage{verbatim} 
\usepackage{fancyhdr}
\usepackage{fancybox}
\usepackage{color}
\usepackage{ulem,bbold}
\usepackage{enumitem}
\usepackage{subfigure}
\usepackage{bbm}
\usepackage{parskip}
\usepackage{cite}

\newcommand{\Comment}[1]{{}}
\definecolor{MyDarkBlue}{rgb}{0.15,0.15,0.45}
\usepackage[linktocpage=true]{hyperref}
\hypersetup{
colorlinks=true,
citecolor=MyDarkBlue,
linkcolor=MyDarkBlue,
urlcolor=MyDarkBlue,
}

\newcommand{\be}{\begin{equation}}
\newcommand{\ee}{\end{equation}}
\newcommand{\bea}{\begin{eqnarray}}
\newcommand{\eea}{\end{eqnarray}}
\newcommand{\beas}{\begin{eqnarray*}}
\newcommand{\eeas}{\end{eqnarray*}}
\newcommand{\nn}{\nonumber}

\numberwithin{equation}{section}

\begin{document}


\begin{center}
{\Large \bf{A St\"uckelberg Approach to Quadratic Curvature \\ \vspace{.2cm} Gravity and its Decoupling Limits}}
\end{center} 
 \vspace{1truecm}
\thispagestyle{empty} \centerline{
{\large {Kurt Hinterbichler}}$^{a,}$\footnote{E-mail: \Comment{\href{mailto:khinterbichler@perimeterinstitute.ca}}{\tt khinterbichler@perimeterinstitute.ca}} 
{\large  {and Mehdi Saravani${}^{a,b,}$}}\footnote{E-mail: \Comment{\href{msaravani@perimeterinstitute.ca}}{\tt msaravani@perimeterinstitute.ca}}
}

\vspace{1cm}

\centerline{{\it ${}^{a}$
Perimeter Institute for Theoretical Physics,}}
\centerline{{\it 31 Caroline St. N, Waterloo, Ontario, N2L 2Y5, Canada}}
 
\vspace{1cm}

\centerline{{\it ${}^{b}$
Department of Physics and Astronomy, University of Waterloo, 
}}
\centerline{{\it Waterloo, ON, N2L 3G1, Canada}}

\begin{abstract} 

Curvature squared terms, when added to the Einstein-Hilbert action and treated non-perturbatively, generically result in the propagation of an extra massive scalar state and an extra massive spin-2 ghost state.  Using the St\"uckelberg trick, we study the high-energy limit in which the mass of the spin-2 state is taken to zero, with strong-coupling scales held fixed.  The St\"uckelberg approach makes transparent the interplay between the ghost graviton and the healthy graviton which allows the theory to evade the usual $\Lambda_3$ strong coupling scale of massive gravity and become renormalizable, at the expense of stability.

\end{abstract}



\section{Introduction}
\parskip=5pt
\normalsize

Einstein gravity, because it is non-renormalizable \cite{Goroff:1985th,vandeVen:1991gw}, is understood as a low energy effective field theory which will be corrected at high energies.  The low energy effect of these corrections is expected to be captured by higher derivative terms added to the Einstein-Hilbert action.  The coefficients of these higher derivative terms are determined by the high energy physics.  Without knowledge of this physics, they are free parameters to be determined by experiment, and the higher derivative terms they come with are only to be used perturbatively to calculate low-energy observables in an expansion in powers of the energy of the observable over the energy scale of new physics \cite{Donoghue:1995cz,Burgess:2003jk}. 

Nevertheless, it has long been of interest to ignore the requirement to treat such terms perturbatively, and to ask what they have to say fully non-perturbatively.  
The motivation is often to gain intuition about the effects Planck physics might produce, or to display various pathologies that a UV completion must ultimately overcome.

The leading higher derivative terms are those with four derivatives.  In four dimensions, of the four possible dimension 4 curvature invariants, $R^2$, $R_{\mu\nu}R^{\mu\nu}$, $R_{\mu\nu\rho\sigma}R^{\mu\nu\rho\sigma}$, $\square R$, two of them, $\square R$ and the Gauss-Bonnet combination $R_{\mu\nu\rho\sigma}R^{\mu\nu\rho\sigma}-4R_{\mu\nu}R^{\mu\nu}+R^2$, are total derivatives, leaving a two dimensional space of possibilities which we may parametrize in terms of $R^2$ and the square of the Weyl tensor,
\be\label{lagm}
S={M_P^2}{}\int d^4x~\sqrt{-g}\left[{1\over 2}R+\frac{1}{12m^2}R^2+\frac{1}{4M^2}C_{\mu\nu\rho\sigma}C^{\mu\nu\rho\sigma}\right].
 \ee
Here $m^2$, $M^2$ are the mass scales of new physics, and $M_P^2$ is the Planck mass scale.

This action has been studied regularly from the non-perturbative viewpoint since the early work \cite{Stelle:1976gc,Stelle:1977ry,Julve:1978xn,Fradkin:1981iu,Fradkin:1981vx,Fradkin:1981hx} (see e.g. the recent work \cite{Lu:2015cqa,Capozziello:2015nga,Cognola:2015uva,Maggiore:2015rma,Alvarez-Gaume:2015rwa,Duplessis:2015xva,Mauro:2015waa,Lu:2015psa,Cai:2015fia}).
In particular, around its Minkowski solution the theory propagates, in addition to the massless graviton, a massive spin-2 degree of freedom with mass square $M^2$ and a massive scalar degree of freedom with mass square $m^2$.  The theory has been argued to be renormalizable, essentially due to the improved $\sim 1/k^4$ behavior of the propagator \cite{Stelle:1976gc}.  The traditional problem, obstructing its status as a complete theory of quantum gravity, is a ghost instability; around the same flat background for which the theory is renormalizable, the kinetic terms for the massless graviton and massive spin-2 have opposite signs, so one of them must always be ghostly.

Here, with the motivations mentioned above, we will continue the study of quadratic gravity in the non-linear regime.  In particular, we will be interested in the high energy limit in which the mass of the spin-2 mode goes to zero while keeping various non-linear scales fixed.  In the case of a pure massive spin-2, this limit is greatly simplified using the St\"uckelberg formulation, in which new fields and gauge symmetries are introduced in order to more easily see the non-linear dynamics of the longitudinal modes of the massive spin-2 \cite{Green:1991pa,Siegel:1993sk,ArkaniHamed:2002sp,Creminelli:2005qk,deRham:2010ik} (see \cite{Hinterbichler:2011tt,deRham:2014zqa} for reviews).  In particular, this formalism has been instrumental in finding fully non-linear theories \cite{deRham:2010kj} free \cite{Hassan:2011hr} from Boulware-Deser modes \cite{Boulware:1973my}.

Since quadratic curvature gravity contains a massive spin-2 mode, it is natural to expect that the St\"uckelberg formulation will simplify the description of its dynamics.
Using the methods of \cite{deRham:2011ca,Paulos:2012xe}, we will see that this is indeed the case, and the St\"uckelberg approach provides a new, clean and transparent way to see many of the known features of quadratic curvature gravity.  In the case of generic interacting massive gravity, there is a natural strong coupling scale $\Lambda_5\sim (M_PM^4)^{1/5}$, and its generalizations in higher and lower dimensions, which sets the scale of unitarity violation for the interactions of longitudinal modes of the massive graviton.  In the case of massive gravity with no Boulware-Deser mode, this scale is raised to $\Lambda_3\sim (M_PM^2)^{1/3}$, and its generalization in other dimensions.  We will see that this higher scale emerges naturally in the St\"uckelberg analysis of quadratic gravity, and that the massive graviton propagated by quadratic gravity has no extra non-linear degrees of freedom.  

The interactions of the longitudinal mode, in the decoupling limit in which the mass is sent to zero with the strong coupling scale held fixed, are described by a cubic galileon.  We find, however, that the non-linear galileon terms are proportional to $D-4$, and hence vanish in the four dimensional case.  In this case, there is no higher strong coupling scale and the theory becomes manifestly renormalizable in the massless limit.  This provides a new way to understand the renormalizability of the theory in four dimensions.  With a single massive graviton, or a ghost-free bi-gravity theory such as those of \cite{Hassan:2011zd,Hinterbichler:2012cn}, it is impossible to raise the strong coupling beyond the $\Lambda_3$ scale \cite{Schwartz:2003vj}. But allowing a relative ghost between the kinetic terms makes this possible, as quadratic curvature gravity demonstrates.   The St\"uckelberg approach makes it easy to see how the ghost and non-ghost graviton interplay and cancel at higher energies in order to render the theory renormalizable.

\section{Second order action and linear degrees of freedom\label{firstsec}}

The St\"uckelberg trick works by restoring the gauge invariance broken by the mass terms of massive fields.  In the case of quadratic curvature gravity, the theory is already diffeomorphism invariant and there is no obvious broken symmetry to restore.  But the theory propagates two gravitons, so we should really think of it as a bi-metric theory, with the massive graviton due to a broken second diffeomorphism invariance.  Thus to apply the St\"uckelberg trick, we must first rewrite the theory in its natural bi-metric form, and then restore the second diffeomorphism. 

Like any higher order theory (with the exception of certain degenerate cases such as \cite{Deser:2009hb}, which we have excluded by demanding the presence of the Einstein-Hilbert term), \eqref{lagm} can be cast into ordinary second order form via the introduction of auxiliary variables.  We start by removing the $R^2$ term through the introduction of a dimension 2 auxiliary scalar $\phi$,
\be\label{lagm2}
S={M_P^2}\int d^4x~\sqrt{-g}\left[{1\over 2}\left(1+{\phi\over 3m^2}\right)R-\frac{1}{12m^2}\phi^2+\frac{1}{4M^2}C_{\mu\nu\rho\sigma}C^{\mu\nu\rho\sigma}\right].
 \ee
The $\phi$ equation of motion fixes $\phi=R$, which upon substitution into \eqref{lagm2} recovers \eqref{lagm}.
We next perform a Weyl field redefinition (which does not effect the Weyl invariant $C^2$ term)
\be g_{\mu\nu}\rightarrow {3m^2\over \phi+3m^2}g_{\mu\nu},\label{weyl1}\ee
followed by a field redefinition 
\be \phi=3m^2\left(e^\psi-1\right),\label{redef1c}\ee
(so that \eqref{weyl1} reads $g_{\mu\nu}\rightarrow e^{-\psi}g_{\mu\nu}$)
which leaves a canonical scalar $\psi$ in Einstein frame
\be S={M_P^2}\int d^4x~\sqrt{-g}\left[{1\over 2}R-{3\over 4}(\partial\psi)^2-{3\over 4}m^2e^{-2\psi}\left(e^\psi-1\right)^2+\frac{1}{4M^2}C_{\mu\nu\rho\sigma}C^{\mu\nu\rho\sigma}\right].\label{lag4}
 \ee
 
Next we want to eliminate the Weyl squared part, which we accomplish through the introduction of a symmetric dimensionless auxiliary tensor field $f_{\mu\nu}$,
\be S={M_P^2}\int d^4x~\sqrt{-g}\left[{1\over 2}R-{3\over 4}(\partial\psi)^2-{3\over 4}m^2e^{-2\psi}\left(e^\psi-1\right)^2 +f^{\mu\nu}G_{\mu\nu}-{1\over 2}M^2\left(f_{\mu\nu}f^{\mu\nu}-f^2\right)\right],\label{lag5b}
 \ee
 where $G_{\mu\nu}$ is the Einstein tensor of $g_{\mu\nu}$, and indices are always moved with $g_{\mu\nu}$.
The $f_{\mu\nu}$ equations of motion can be solved to give $f_{\mu\nu}={1\over M^2}\left(R_{\mu\nu}-{1\over 6}Rg_{\mu\nu}\right)$, which when plugged into \eqref{lag5b} recovers \eqref{lag4}.  The theory is now manifestly second order.

This second order action is the easiest starting point from which to see the linear spectrum of fluctuations at the Lagrangian level.  Expanding to second order in fluctuations around the background $g_{\mu\nu}=\eta_{\mu\nu}$, $f_{\mu\nu}=0$, $\psi=0$, with the metric fluctuation defined as $g_{\mu\nu}=\eta_{\mu\nu}+h_{\mu\nu}$, we have the flat space linear action
\be S_2={M_P^2}\int d^4x~-{3\over 4}\left((\partial\psi)^2+m^2\psi^2\right) +{1\over 8}h^{\mu\nu}\left({\mathcal E}h\right)_{\mu\nu}-{1\over 2}f^{\mu\nu}\left({\mathcal E}h\right)_{\mu\nu}-{1\over 2}M^2\left(f_{\mu\nu}f^{\mu\nu}-f^2\right),\label{lag5}
 \ee
where $\left({\mathcal E}h\right)_{\mu\nu}\equiv\square h_{\mu\nu}-\eta_{\mu\nu}\square h-2\partial_{(\mu}\partial^\rho h_{\nu)\rho}+\partial_\mu\partial_\nu h+\eta_{\mu\nu}\partial^\rho\partial^\sigma h_{\rho\sigma}$ is the standard graviton kinetic operator.  We may diagonalize the tensor kinetic terms with the field redefinition 
\be h_{\mu\nu}=2\left(h_{\mu\nu}'+f_{\mu\nu}\right),\ee
after which we have
\be S={M_P^2}\int d^4x~-{3\over 4}\left((\partial\psi)^2-m^2\psi^2\right) +{1\over 2}h'^{\mu\nu}\left({\mathcal E}h'\right)_{\mu\nu}-{1\over 2}f^{\mu\nu}\left({\mathcal E}f\right)_{\mu\nu}-{1\over 2}M^2\left(f_{\mu\nu}f^{\mu\nu}-f^2\right),\label{lag5}
 \ee
with the (in)famous relative minus sign between the two tensor modes.  The degrees of freedom are:
\begin{enumerate}
\item a massive scalar field $ \psi$, with mass squared ${m^2}$,
\item a massless spin-2 field $ h_{ab}'$,
\item a massive (ghost) spin-2 field $ f_{ab}$, with mass squared $M^2$.
\end{enumerate}
We can make the massive spin-2 healthy, at the expense of making the massless spin-2 and scalar ghostly, by flipping the overall sign of the action, but we cannot remove all the instabilities\footnote{Some approaches toward the ghost problem are to break Lorentz invariance \cite{Chen:2013aha}, sacrifice unitarity \cite{Hawking:2001yt}, try to quantize in a non-standard fashion \cite{Mannheim:2011ds}, introduce non-locality \cite{Modesto:2011kw,Biswas:2011ar,Conroy:2014eja}, argue that the ghost is not in the physical spectrum, \cite{Holdom:2015kbf} or try to argue that something cuts off the infinite phase space integral in the decay rate of the vacuum, making the vacuum long-lived enough to be acceptable \cite{Cline:2003gs}.}.

\section{St\"uckelberg}

In this section we will generalize to $D$ dimensions in order to illustrate cancellations that occur for $D=4$.  The scalar $\psi$ plays no role in what follows and merely comes for the ride, so we will temporarily drop it, starting with the fourth order action containing only the massless and massive spin-2 degrees of freedom,
\be  S={M_P^{D-2}}\int d^Dx~\sqrt{-g}\left[{1\over 2}R+{1\over 2 M^2}\left(R_{\mu\nu}R^{\mu\nu}-{D\over 4(D-1)}R^2\right)\right].\label{lagD}\ee
The second order form is 
\be S={M_P^{D-2}}\int d^Dx~\sqrt{-g}\left[{1\over 2}R +f^{\mu\nu}G_{\mu\nu}-{1\over 2}M^2\left(f_{\mu\nu}f^{\mu\nu}-f^2\right)\right].\label{lag5D}
 \ee
After using the $f_{\mu\nu}$ equations of motion to set $f_{\mu\nu}={1\over M^2}\left(R_{\mu\nu}-{1\over 2(D-1)}Rg_{\mu\nu}\right)$, we recover \eqref{lagD}.

The action \eqref{lag5b} has ordinary diffeomorphism invariance, under which $f_{\mu\nu}$ (and $\psi$) transforms as an ordinary tensor.  But it is really a two-tensor theory propagating a massive spin-2 mode.  A massive spin-2 propagates vector and scalar longitudinal modes, so following \cite{ArkaniHamed:2002sp}, we should introduce a second diffeomorphism symmetry and a U(1) in order to make all the physics manifest.  We do this through the St\"uckelberg replacement 
\be f_{\mu\nu}\rightarrow f_{\mu\nu}+\nabla_\mu \tilde V_\nu+\nabla_\nu \tilde V_\mu,\ \ \ \tilde V_\mu=V_\mu+\partial_\mu\pi.\label{stukrepm}\ee
We have introduced two new fields $V_\mu$ and $\pi$, along with two new gauge symmetries with gauge parameters $\Lambda_\mu$ and $\Lambda$,
\bea  &&\delta f_{\mu\nu}=\nabla_\mu \Lambda_\nu+ \nabla_\nu \Lambda_\mu,\ \ \ \ \delta V_\mu=-\Lambda_\mu+\partial_\mu\Lambda, \ \ \  \delta\pi= \Lambda  \ .\label{secondgfullt}
\eea
The action \eqref{lag5D} now takes the form
\bea S&=&{M_P^{D-2}}\int d^Dx~\sqrt{-g}\bigg[{1\over 2}R+f^{\mu\nu}G_{\mu\nu}-{1\over 2}M^2\left(f_{\mu\nu}f^{\mu\nu}-f^2\right)-{1\over 2}M^2F_{\mu\nu}^2\nn\\
&&+2M^2R_{\mu\nu}\tilde V^\mu \tilde V^\nu -2M^2 f^{\mu\nu}\left(\nabla_\mu \tilde V_\nu-g_{\mu\nu} \nabla\cdot \tilde V\right)\bigg], \label{lag6}
 \eea
where $F_{\mu\nu}=\nabla_\mu \tilde V_\nu-\nabla_\nu \tilde V_\mu=\nabla_\mu  V_\nu-\nabla_\nu  V_\mu$ is the Maxwell field strength of $V_{\mu}$.  All covariant derivatives and index movements are with respect to $g_{\mu\nu}$.

The full non-linear degree of freedom counting is now manifest \cite{deRham:2011ca,Paulos:2012xe}.  The theory has been cast into second order form with purely first class gauge symmetries, so the degree of freedom count is $({\rm number\ of\ fields})-2({\rm number\ of\ gauge\ symmetries})$.  The gauge strikes twice because one field will be a Lagrange multiplier which enforces the gauge constraint.  For example, in $D=4$ we have 25 fields (two symmetric tensors with 10 components each, one vector with 4 components, and one scalar) and 9 gauge symmetries (two diffeomorphisms with 4 components each, and a $U(1)$), which leaves $25-2\cdot 9=7$ degrees of freedom, the correct number for a massless graviton and a massive graviton.  Thus there is no Boulware-Deser like \cite{Boulware:1973my} extra degree of freedom associated with the massive spin-2.

\section{Decoupling limit}

After canonically normalizing the fields (note that the kinetic term for $\pi$ comes from mixing with $f_{\mu\nu}$),
\be (h_{\mu\nu},f_{\mu\nu}) \sim {1\over M_P^{{D\over 2}-1}}(\hat h_{\mu\nu},\hat f_{\mu\nu}) ,\ \ \ V_\mu\sim {1\over M_P^{{D\over 2}-1}M}\hat V_\mu,\ \ \ \pi\sim {1\over M_P^{{D\over 2}-1}M^2}\hat\pi,\ee
we can read off the strong coupling scale from any given interaction term.
The lowest possible scales are those coming from $\pi$ self-interactions or interactions with one $V$ and the rest $\pi$, but these are not present in \eqref{lag6}.  The lowest scale present in \eqref{lag6} is 
\be\label{scale}\Lambda_{D+2\over D-2}= \left(M^{4\over D-2}M_P\right)^{D-2\over D+2},\ee
coming from self interactions with one $h$ or $f$ and the rest $\pi$'s.  
We will be interested in taking the decoupling limit
\be M\rightarrow 0,\ \ \ \ M_P\rightarrow \infty,\ \ \ \ \Lambda_{D+2\over D-2}\ {\rm fixed}.\ee
This is a high energy limit in which the massive graviton is becoming massless with the leading strong coupling scale held fixed.

The action in this limit reduces to the flat space action
\bea S={M_P^{D-2}}\int d^Dx~\bigg[&&{1\over 8}h^{\mu\nu}\left({\mathcal E}h\right)_{\mu\nu}-{1\over 2}f^{\mu\nu}\left({\mathcal E}h\right)_{\mu\nu}-{1\over 2}M^2F_{\mu\nu}^2-2M^2 f^{\mu\nu}\left(\partial_\mu \partial_\nu\pi-\eta_{\mu\nu} \square\pi\right) \nn\\
&&+ 2M^2 R_{\mu\nu}^{\rm L}(h)\partial^\mu\pi\partial^\nu\pi \bigg],
 \label{laglambda3}
 \eea
where $R_{\mu\nu}^{\rm L}(h)$ is the linearized Ricci tensor.  The gauge symmetries in the decoupling limit reduce to their linear versions,
\bea \delta h_{\mu\nu}&=& \partial_\mu\xi_\nu+\partial_\nu\xi_\nu ,\nn\\
\delta f_{\mu\nu}&=&  \partial_\mu\Lambda_\nu+\partial_\nu\Lambda_\mu ,\nn\\
\delta V_{\mu}&=&  \partial_\mu\Lambda ,\nn\\
\delta \pi_{}&=&0,
\eea
with $\xi^\mu$ the diffeomorphism parameter.  It is easy to see that \eqref{laglambda3} is invariant under these.

We can decouple the scalar and diagonalize the kinetic terms by making the field redefinition
\bea && h_{\mu\nu}\rightarrow 2\left(h'_{\mu\nu}+f'_{\mu\nu}\right)-{4\over D-2}M^2\eta_{\mu\nu}\pi,\nn\\
&& f_{\mu\nu}\rightarrow f'_{\mu\nu}-{2\over D-2}M^2\eta_{\mu\nu}\pi-2M^2\left[\partial_\mu\pi\partial_\nu\pi-{1\over D-2}(\partial\pi)^2\eta_{\mu\nu}\right],
\eea
after which the action becomes
\bea S={M_P^{D-2}}\int d^Dx~\bigg[&&{1\over 2}h'^{\mu\nu}\left({\mathcal E}h'\right)_{\mu\nu}-{1\over 2}f'^{\mu\nu}\left({\mathcal E}f'\right)_{\mu\nu}-{1\over 2}M^2F_{\mu\nu}^2+{2(D-1)\over D-2}M^4(\partial\pi)^2 \nn\\
&&-{2M^4 (D-4)\over D-2}(\partial\pi)^2\square\pi\bigg]. \label{laglambda3b}
 \eea
 
The only interaction is the final term in \eqref{laglambda3b}, which is a cubic galileon interaction\footnote{The galileon has a well-known global symmetry $\pi\rightarrow \pi+c+b_\mu x^\mu$, for constants $c,b_\mu$ where $x^\mu$ is the spacetime coordinate, stemming from the fact that $\pi$ always appears with two derivatives in \eqref{stukrepm}.  The $D=3$ case is special in that the galileon interaction in \eqref{laglambda3} has an enhanced shift symmetry \cite{Hinterbichler:2015pqa}.  It is not yet clear what the gravitational origin of this might be.}\cite{Luty:2003vm,Nicolis:2008in}.  For $D\not=4$, this describes the non-linear high-energy dynamics of the longitudinal mode of the massive graviton.  It is straightforward to see that the 4 particle amplitude for $\pi$ scattering is non-vanishing and violates perturbative unitarity at the scale $\Lambda_{D+2\over D-2}$, so the theory is perturbatively non-renormalizable at this intermediate scale.  Any physics lost in the decoupling limit cannot enter until a higher scale parametrically suppressed by $M_P$, so as long as $M\ll M_P$, so that the decoupling limit makes sense, there is a regime in which perturbative unitarity is violated.  For $D=3$, \eqref{lagD} reduces to the case of new massive gravity \cite{Bergshoeff:2009hq} (studied using the methods here in \cite{deRham:2011ca}), which was argued to be non-renormalizable in \cite{Muneyuki:2012ur}.  Quadratic gravity for $D>4$ has also been argued to be non-renormalizable \cite{Muneyuki:2013aba}.

For $D=4$, however, the galileon interaction term vanishes, a signal that the true strong coupling scale is higher.  For this case, we must search for non-trivial operators at higher scales.

\section{Massless limit}

As we will see now, in $D=4$ there is in fact no higher intermediate scale for which there are non-trivial interaction terms, and thus there is no obstruction to taking a straight $M\rightarrow 0$ limit with $M_P$ held fixed.

To see this, fix $D=4$ in \eqref{lag6} and make the field redefinition
\be f_{\mu\nu}\rightarrow f'_{\mu\nu}+{1\over 2}g_{\mu\nu}-2M^2\left[\tilde V_\mu \tilde V_\nu-{1\over 2}g_{\mu\nu}\tilde V^2\right],
\ee
after which the action has the following finite and smooth limit as $M^2\rightarrow 0$ with the canonically normalized fields held fixed,
\bea S={M_P^{2}}\int d^4x~\sqrt{-g}\bigg[&-&{1\over 2}M^2F_{\mu\nu}^2 +3M^4(\partial\pi)^2\nn\\ 
&+&f'^{\mu\nu}\left(G_{\mu\nu}-2M^2\left(\nabla_\mu\nabla_\nu\pi-g_{\mu\nu}\square\pi\right)+2M^4\left(\nabla_\mu\pi\nabla_\nu\pi+{1\over 2}g_{\mu\nu}(\partial\pi)^2\right)\right)\bigg] .\label{lag8} \nn\\
 \eea
The gauge symmetries in the massless limit reduce to ordinary diffeomorphisms for $g_{\mu\nu}$, $f_{\mu\nu}$, $V_\mu$ and $\pi$ along with the massless limit of the second set of symmetries \eqref{secondgfullt} expressed in terms of $f'_{\mu\nu}$,
\bea \delta f_{\mu\nu}&=& \nabla_\mu\Lambda_\nu+\nabla_\nu\Lambda_\mu-2M^2\left(\nabla_\mu\pi\,\Lambda_\nu+\nabla_\nu\pi\,\Lambda_\mu-g_{\mu\nu}\nabla_\rho\pi\,\Lambda^\rho\right) ,\nn\\
\delta V_{\mu}&=&  \partial_\mu\Lambda ,\nn\\
\delta \pi&=&0. \label{gaugemassfp}
\eea
The action \eqref{lag8} is invariant under these transformations.

Noting that $G_{\mu\nu}\left[e^{2M^2\pi}g_{\mu\nu}\right]=G_{\mu\nu}-2M^2\left(\nabla_\mu\nabla_\nu\pi-g_{\mu\nu}\square\pi\right)+2M^4\left(\nabla_\mu\pi\nabla_\nu\pi+{1\over 2}g_{\mu\nu}(\partial\pi)^2\right)$, we can simplify \eqref{lag8} by a making a conformal transformation 
\be g_{\mu\nu}\rightarrow e^{-2M^2\pi}g_{\mu\nu},\label{conft2s}\ee
after which it becomes
\bea S={M_P^{2}}\int d^4x~\sqrt{-g}\left[f'^{\mu\nu}G_{\mu\nu}-{1\over 2}M^2F_{\mu\nu}^2+3M^4e^{-2M^2\pi}(\partial\pi)^2\right], \label{lag9} 
 \eea
 and the gauge symmetry \eqref{gaugemassfp} becomes (taking $\Lambda^\mu$ to be independent of the metric)
 \bea \delta f_{\mu\nu}&=& \nabla_\mu\Lambda_\nu+\nabla_\nu\Lambda_\mu,\nn\\
\delta V_{\mu}&=&  \partial_\mu\Lambda ,\nn\\
\delta \pi&=&0.
\eea
This action describes the high energy dynamics of quadratic gravity in four dimensions.  

If quadratic gravity is renormalizable, there should be no non-trivial non-renormalizable operators present at any scale, even $M_P$.  We will now argue that this is indeed the case for \eqref{lag9}.  Upon expanding\footnote{Note that in this massless limit there is now a moduli space of solutions $f'_{\mu\nu}=c \eta_{\mu\nu}$, $g_{\mu\nu}=\eta_{\mu\nu}$ for constant $c$, which is not present away from the massless limit.  To keep the solution which exists away from the massless limit, we take $c={1\over 2}$ corresponding to the background where $f_{\mu\nu}=0$.} $g_{\mu\nu}=\eta_{\mu\nu}+h_{\mu\nu}$, $f'_{\mu\nu}={1\over 2}\eta_{\mu\nu}+\delta f_{\mu\nu}$ and then diagonalizing the two graviton kinetic terms with the redefinition,
\be h_{\mu\nu}=\tilde h_{\mu\nu}+\tilde f_{\mu\nu},\ \ \ \delta f_{\mu\nu}=\tilde f_{\mu\nu}-{1\over 2} \tilde h_{\mu\nu}, \ee 
the action expanded around flat space reads
\bea S={M_P^{2}}\int d^4x~\bigg[&& {3\over 8}\tilde h^{\mu\nu}\left({\mathcal E}\tilde h\right)_{\mu\nu}-{3\over 8}\tilde f^{\mu\nu}\left({\mathcal E}\tilde f\right)_{\mu\nu} -{1\over 2}M^2F_{\mu\nu}^2+3M^4e^{-2M^2\pi}(\partial\pi)^2\nn\\ 
&&+\left(\tilde f_{\mu\nu}-{1\over 2}\tilde h_{\mu\nu}\right)\sqrt{-g}G^{(\geq 2)\mu\nu}\left[\tilde h+\tilde f\right]+{\cal L}_{V,\pi}^{(\geq 1)}\left[\tilde h+\tilde f,V,\pi\right]\bigg]. \nn\\ \label{schemlag1h}
\eea
Here $\sqrt{-g}G^{(\geq 2)\mu\nu}\left[h\right]$ stands for the terms of order $h^2$ and higher obtained from expanding the Einstein tensor and metric determinant, and ${\cal L}_{V,\pi}^{(\geq 1)}\left[h,V,\pi\right]$ the terms of order $h$ and higher obtained from expanding the minimally coupled $V$ and $\pi$ Lagrangians.

We see that the scalar and vector couple only to the combination $(\tilde h+\tilde f)_{\mu\nu}$.  Since $\tilde h$ and $\tilde f$ have equal and opposite propagators, and equal couplings to $V,\pi$, we can see that there will be a cancellation in pairs among all Feynman diagrams with external $V,\pi$ lines.  For each diagram with an internal $\tilde h$, there is an equal and opposite one in which the internal $\tilde h$ is replaced by an internal $\tilde f$.  This is the mechanism by which the theory becomes renormalizable; the bad high energy behavior of the graviton cancels against the bad high energy behavior of the ghost.

This leads us to suspect that the action \eqref{lag9} is in fact a free action in disguise, as we will now argue.  
The key observation is that the kinetic terms for $\tilde h,\tilde f$ are invariant under an internal $SO(1,1)$ symmetry, so making the following field redefinition, depending on some parameter $\alpha$,
\bea \left(\begin{array}{c} \tilde h_{\mu\nu} \\  \tilde f_{\mu\nu}\end{array}\right)=\left(\begin{array}{cc}\cosh\alpha & \sinh \alpha \\ \sinh\alpha & \cosh\alpha\end{array}\right)\left(\begin{array}{c}\tilde h_{\mu\nu}^{(\alpha)} \\ \tilde f_{\mu\nu}^{(\alpha)}\end{array}\right) ,\label{alpharedef1}
\eea
the kinetic terms remain invariant and the action \eqref{schemlag1h} becomes
\bea   S={M_P^{2}}\int d^4x~\bigg[&& {3\over 8}\tilde h^{(\alpha)\mu\nu}\left({\mathcal E}\tilde h^{(\alpha)}\right)_{\mu\nu}-{3\over 8}\tilde f^{{(\alpha)}\mu\nu}\left({\mathcal E}\tilde f^{(\alpha)}\right)_{\mu\nu}-{1\over 2}M^2F_{\mu\nu}^2+3M^4e^{-2M^2\pi}(\partial\pi)^2 \nn\\
&&+\left(\tilde f^{}-{1\over 2}\tilde h ^{} \right)_{\mu\nu}\sqrt{-g}G^{(\geq 2)\mu\nu}\left[e^\alpha\left(\tilde h^{(\alpha)}+\tilde f^{(\alpha)}\right)\right]\nn\\ 
&&+{\cal L}_{V,\pi}^{(\geq 1)}\left[e^\alpha\left(\tilde h^{(\alpha)}+\tilde f^{(\alpha)}\right),V,\pi\right]\bigg]. \label{schemlag2m}
\eea
Now take the limit $\alpha\rightarrow-\infty$.  All the gravitational interactions, i.e. the final two lines of \eqref{schemlag2m}, scale away, and we are left with the flat-space action of the first line
\be S={M_P^{2}}\int d^4x~{3\over 8}\tilde h^{(\alpha)\mu\nu}\left({\mathcal E}\tilde h^{(\alpha)}\right)_{\mu\nu}-{3\over 8}\tilde f^{{(\alpha)}\mu\nu}\left({\mathcal E}\tilde f^{(\alpha)}\right)_{\mu\nu}-{1\over 2}M^2F_{\mu\nu}^2+3M^4e^{-2M^2\pi}(\partial\pi)^2 .\ee   
This is a completely free theory (the scalar self-interactions can be absorbed with a field re-definition $\pi\rightarrow-{1\over M^2}\log (M^2\pi)$ ), thus the high energy dynamics of the theory is trivial, illustrating why the theory is renormalizable.  We see clearly the role that the ghost graviton plays in making this work.  At high energies, the ghost graviton interactions cancel precisely the standard gravitational interactions, rendering the theory asymptotically free.

If we bring back the scalar field $\psi$ from section \ref{firstsec}, keeping its mass $m^2$ fixed as we scale $M\rightarrow 0$, and remembering the conformal transformation \eqref{conft2s}, we find, after scaling 
\be \pi\rightarrow {\pi\over 2M^2},\label{redef2s}\ee
that the $\alpha\rightarrow-\infty$ limiting action becomes the flat space action
\be S={M_P^{2}}\int d^4x~\left[{3\over 4}e^{-\pi}(\partial\pi)^2-{3\over 4}e^{-\pi}(\partial\psi)^2-{3\over 4}m^2e^{-2(\psi+\pi)}\left(e^\psi-1\right)^2\right] ,\label{sigmodels}\ee
in addition to the free vector and two free gravitons.  
Now we see a relative ghost sign between the scalar $\psi$ coming from the original $R^2$ term and the scalar $\pi$ coming from the longitudinal mode of the massive graviton.  The potential in \eqref{sigmodels} has a moduli space of vacua along the line $\psi=0$, $\pi=c$ parametrized by the constant $c$.  The $\psi$ field has mass $m_\psi^2=m^2e^{-2c}$ along this line whereas $\pi$ remains massless.  

\eqref{sigmodels} is a sigma model with two-dimensional Minkowski target space in a Milne slicing.  We can canonicalize the kinetic terms by going to flat field space coordinates via the field redefinition
\be  \pi=-\log\left(\tilde\pi^2-\tilde \psi^2\right),\ \ \  \psi=\log\left(\tilde \pi+\tilde\psi\over \tilde \pi-\tilde\psi\right),\ee
after which \eqref{sigmodels} becomes
\be S=3{M_P^{2}}\int d^4x~(\partial\tilde\pi)^2-(\partial\tilde\psi)^2+m^2\tilde\psi^2\left(\tilde \pi-\tilde\psi\right)^2 .\label{sigmodels2}\ee
This allows us to analytically continue the range of field space; the region covered by $(\pi,\psi)$ corresponds to the region $\tilde\pi^2>\tilde\psi^2$, $\tilde\pi>0$.  The moduli line of vacua $\psi=0$ corresponds to the line $\tilde \psi=0$, with our original vacuum $(\pi=0,\psi=0)$ corresponding to $(\tilde \pi=1,\tilde\psi=0)$, and the point $c\rightarrow \infty$ where the fields become massless corresponding to the origin $(\tilde \pi=0,\tilde\psi=0)$.  In addition, there is a new line of vacua given by $\tilde\pi=\tilde\psi$, which is not covered by the original $\pi,\psi$ coordinates.

The action \eqref{sigmodels2} contains only renormalizable interactions, with mass terms of order $\sim m^2$ and couplings of order $\sim m^2/M_P^2$.  Thus, even with $R^2$ terms, we see explicitly the absence of strong coupling scales in the quadratic gravity decoupling limit, reflecting the renormalizability of the theory.

As with the gravitons, the scalar kinetic terms in \eqref{sigmodels2} have an internal $SO(1,1)$ symmetry.  Our original vacuum at $(\tilde \pi=1,\tilde\psi=0)$ is not invariant under this action, but we can simplify the description of the S-matrix about the massless $(\tilde \pi=0,\tilde\psi=0)$ vacuum by making the following field redefinition,
\bea \left(\begin{array}{c} \tilde \psi \\  \tilde \pi\end{array}\right)=\left(\begin{array}{cc}\cosh\alpha & \sinh \alpha \\ \sinh\alpha & \cosh\alpha\end{array}\right)\left(\begin{array}{c}\tilde \psi^{(\alpha)} \\ \tilde \pi^{(\alpha)}\end{array}\right). \label{alpharedef2}
\eea
The kinetic terms remain invariant and the action \eqref{sigmodels2} becomes, in the limit $\alpha\rightarrow \infty$, 
\bea S=3{M_P^{2}}\int d^4x ~(\partial\tilde\pi^{(\alpha)})^2-(\partial\tilde\psi^{(\alpha)})^2 -{1\over 4}m^2\left(\tilde \pi^{(\alpha)2}-\tilde\psi^{(\alpha)2}\right)^2.\label{sigmodels3}
\eea

Finally, consider coupling some additional matter field to the original metric $g_{\mu\nu}$ of \eqref{lagm}.  Tracing back through the conformal transformations \eqref{weyl1}, \eqref{conft2s}, and the field redefinitions \eqref{redef1c}, \eqref{redef2s} the matter fields couples to the combination $e^{-(\pi+\psi)}g_{\mu\nu}=\left(\tilde\pi-\tilde\psi\right)^2\left(\eta_{\mu\nu}+\tilde h_{\mu\nu}+\tilde f_{\mu\nu}\right)$, which scales away after the redefinitions \eqref{alpharedef1}, \eqref{alpharedef2} in the corresponding limits.  Thus we see that the addition of a (renormalizable) matter sector does not spoil the renormalizability of quadratic gravity.

\section{Summary and conclusions}

Using the St\"uckelberg trick, we have studied the non-linear dynamics of quadratic curvature gravity in the limit in which the mass of the second graviton goes to zero.  The St\"uckelberg fields account for the longitudinal modes of the massive graviton, and restore a second diffeomorphism invariance associated with the second graviton.  In dimensions $D\not= 4$, the non-linear dynamics are described by a cubic galileon term, becoming strongly coupled at the scale $\Lambda_{D+2\over D-2}$ associated with a non-linear massive graviton with no Boulware-Deser mode.  In $D=4$, the galileon term vanishes, and the theory never becomes strongly coupled, becoming a renormalizable theory in the massless limit.  The ghostly second graviton is crucial in making this happen, and the St\"uckelberg trick makes transparent the mechanism by which it works.

Though we have studied only quadratic curvature gravity, there is no obstruction in principle to applying this kind of St\"uckelberg analysis to all varieties of higher-order gravitational Lagrangians, and simplifying the non-linear dynamics of the theory in the high-energy limit (as in e.g. \cite{Paulos:2012xe}).  It need only be ensured that the St\"uckelberg fields faithfully represent the true degrees of freedom of the theory.


\bigskip
{\bf Acknowledgements}: 
The authors would like to thank Niayesh Afshordi for helpful discussions. Research at Perimeter Institute is supported by the Government of Canada through Industry Canada and by the Province of Ontario through the Ministry of Economic Development and Innovation. This work was made possible in part through the support of a grant from the John Templeton Foundation. The opinions expressed in this publication are those of the authors and do not necessarily reflect the views of the John Templeton Foundation (KH).

\bibliographystyle{utphys}
\addcontentsline{toc}{section}{References}
\bibliography{curvaturesquaregravityarxiv}

\providecommand{\href}[2]{#2}\begingroup\raggedright\begin{thebibliography}{10}

\bibitem{Goroff:1985th}
M.~H. Goroff and A.~Sagnotti, ``{The Ultraviolet Behavior of Einstein
  Gravity},''
\href{http://dx.doi.org/10.1016/0550-3213(86)90193-8}{{\em Nucl. Phys.}
  {\bfseries B266} (1986) 709}.

\bibitem{vandeVen:1991gw}
A.~E.~M. van~de Ven, ``{Two loop quantum gravity},''
\href{http://dx.doi.org/10.1016/0550-3213(92)90011-Y}{{\em Nucl. Phys.}
  {\bfseries B378} (1992) 309--366}.

\bibitem{Donoghue:1995cz}
J.~F. Donoghue, ``{Introduction to the effective field theory description of
  gravity},''
\href{http://arxiv.org/abs/gr-qc/9512024}{{\ttfamily arXiv:gr-qc/9512024
  [gr-qc]}}.

\bibitem{Burgess:2003jk}
C.~Burgess, ``{Quantum gravity in everyday life: General relativity as an
  effective field theory},'' \href{http://dx.doi.org/10.12942/lrr-2004-5}{{\em
  Living Rev.Rel.} {\bfseries 7} (2004) 5--56},
\href{http://arxiv.org/abs/gr-qc/0311082}{{\ttfamily arXiv:gr-qc/0311082
  [gr-qc]}}.

\bibitem{Stelle:1976gc}
K.~Stelle, ``{Renormalization of Higher Derivative Quantum Gravity},''
\href{http://dx.doi.org/10.1103/PhysRevD.16.953}{{\em Phys.Rev.} {\bfseries
  D16} (1977) 953--969}.

\bibitem{Stelle:1977ry}
K.~S. Stelle, ``{Classical Gravity with Higher Derivatives},''
\href{http://dx.doi.org/10.1007/BF00760427}{{\em Gen.Rel.Grav.} {\bfseries 9}
  (1978) 353--371}.

\bibitem{Julve:1978xn}
J.~Julve and M.~Tonin, ``{Quantum Gravity with Higher Derivative Terms},''
\href{http://dx.doi.org/10.1007/BF02748637}{{\em Nuovo Cim.} {\bfseries B46}
  (1978) 137--152}.

\bibitem{Fradkin:1981iu}
E.~S. Fradkin and A.~A. Tseytlin, ``{Renormalizable asymptotically free quantum
  theory of gravity},''
\href{http://dx.doi.org/10.1016/0550-3213(82)90444-8}{{\em Nucl. Phys.}
  {\bfseries B201} (1982) 469--491}.

\bibitem{Fradkin:1981vx}
E.~S. Fradkin and A.~A. Tseytlin, ``{Asymptotic Freedom in Renormalizable
  Gravity and Supergravity},'' in {\em {In *Moscow 1981, Proceedings, Quantum
  Gravity*, 29-45}}.
\newblock
1981.
\newblock

\bibitem{Fradkin:1981hx}
E.~S. Fradkin and A.~A. Tseytlin, ``{Renormalizable Asymptotically Free Quantum
  Theory of Gravity},''
\href{http://dx.doi.org/10.1016/0370-2693(81)90702-4}{{\em Phys. Lett.}
  {\bfseries B104} (1981) 377--381}.

\bibitem{Lu:2015cqa}
H.~Lu, A.~Perkins, C.~Pope, and K.~Stelle, ``{Black Holes in Higher-Derivative
  Gravity},'' \href{http://dx.doi.org/10.1103/PhysRevLett.114.171601}{{\em
  Phys.Rev.Lett.} {\bfseries 114} no.~17, (2015) 171601},
\href{http://arxiv.org/abs/1502.01028}{{\ttfamily arXiv:1502.01028 [hep-th]}}.

\bibitem{Capozziello:2015nga}
S.~Capozziello and A.~Stabile, ``{Gravitational waves in fourth order
  gravity},''
\href{http://dx.doi.org/10.1007/s10509-015-2425-1}{{\em Astrophys.Space Sci.}
  {\bfseries 358} no.~2, (2015) 27}.

\bibitem{Cognola:2015uva}
G.~Cognola, M.~Rinaldi, and L.~Vanzo, ``{Scale-invariant rotating black holes
  in quadratic gravity},''
\href{http://arxiv.org/abs/1506.07096}{{\ttfamily arXiv:1506.07096 [gr-qc]}}.

\bibitem{Maggiore:2015rma}
M.~Maggiore, ``{Dark energy and dimensional transmutation in $R^2$ gravity},''
\href{http://arxiv.org/abs/1506.06217}{{\ttfamily arXiv:1506.06217 [hep-th]}}.

\bibitem{Alvarez-Gaume:2015rwa}
L.~Alvarez-Gaume, A.~Kehagias, C.~Kounnas, D.~Lust, and A.~Riotto, ``{Aspects
  of Quadratic Gravity},''
\href{http://arxiv.org/abs/1505.07657}{{\ttfamily arXiv:1505.07657 [hep-th]}}.

\bibitem{Duplessis:2015xva}
F.~Duplessis and D.~A. Easson, ``{Exotica ex nihilo: Traversable wormholes and
  non-singular black holes from the vacuum of quadratic gravity},''
\href{http://arxiv.org/abs/1506.00988}{{\ttfamily arXiv:1506.00988 [gr-qc]}}.

\bibitem{Mauro:2015waa}
S.~Mauro, R.~Balbinot, A.~Fabbri, and I.~L. Shapiro, ``{Fourth derivative
  gravity in the auxiliary fields representation and application to the black
  hole stability},''
\href{http://arxiv.org/abs/1504.06756}{{\ttfamily arXiv:1504.06756 [gr-qc]}}.

\bibitem{Lu:2015psa}
H.~LŸ, A.~Perkins, C.~N. Pope, and K.~S. Stelle, ``{Spherically Symmetric
  Solutions in Higher-Derivative Gravity},''
\href{http://arxiv.org/abs/1508.00010}{{\ttfamily arXiv:1508.00010 [hep-th]}}.

\bibitem{Cai:2015fia}
Y.-F. Cai, G.~Cheng, J.~Liu, and H.~Zhang, ``{Features and stability analysis
  of non-Schwarzschild black hole in quadratic gravity},''
\href{http://arxiv.org/abs/1508.04776}{{\ttfamily arXiv:1508.04776 [hep-th]}}.

\bibitem{Green:1991pa}
M.~B. Green and C.~B. Thorn, ``{Continuing between closed and open strings},''
\href{http://dx.doi.org/10.1016/0550-3213(91)90022-P}{{\em Nucl.Phys.}
  {\bfseries B367} (1991) 462--484}.

\bibitem{Siegel:1993sk}
W.~Siegel, ``{Hidden gravity in open string field theory},''
  \href{http://dx.doi.org/10.1103/PhysRevD.49.4144}{{\em Phys.Rev.} {\bfseries
  D49} (1994) 4144--4153},
\href{http://arxiv.org/abs/hep-th/9312117}{{\ttfamily arXiv:hep-th/9312117
  [hep-th]}}.

\bibitem{ArkaniHamed:2002sp}
N.~Arkani-Hamed, H.~Georgi, and M.~D. Schwartz, ``{Effective field theory for
  massive gravitons and gravity in theory space},''
  \href{http://dx.doi.org/10.1016/S0003-4916(03)00068-X}{{\em Annals Phys.}
  {\bfseries 305} (2003) 96--118},
\href{http://arxiv.org/abs/hep-th/0210184}{{\ttfamily arXiv:hep-th/0210184
  [hep-th]}}.

\bibitem{Creminelli:2005qk}
P.~Creminelli, A.~Nicolis, M.~Papucci, and E.~Trincherini, ``{Ghosts in massive
  gravity},'' \href{http://dx.doi.org/10.1088/1126-6708/2005/09/003}{{\em JHEP}
  {\bfseries 0509} (2005) 003},
\href{http://arxiv.org/abs/hep-th/0505147}{{\ttfamily arXiv:hep-th/0505147
  [hep-th]}}.

\bibitem{deRham:2010ik}
C.~de~Rham and G.~Gabadadze, ``{Generalization of the Fierz-Pauli Action},''
  \href{http://dx.doi.org/10.1103/PhysRevD.82.044020}{{\em Phys.Rev.}
  {\bfseries D82} (2010) 044020},
\href{http://arxiv.org/abs/1007.0443}{{\ttfamily arXiv:1007.0443 [hep-th]}}.

\bibitem{Hinterbichler:2011tt}
K.~Hinterbichler, ``{Theoretical Aspects of Massive Gravity},''
  \href{http://dx.doi.org/10.1103/RevModPhys.84.671}{{\em Rev.Mod.Phys.}
  {\bfseries 84} (2012) 671--710},
\href{http://arxiv.org/abs/1105.3735}{{\ttfamily arXiv:1105.3735 [hep-th]}}.

\bibitem{deRham:2014zqa}
C.~de~Rham, ``{Massive Gravity},''
  \href{http://dx.doi.org/10.12942/lrr-2014-7}{{\em Living Rev.Rel.} {\bfseries
  17} (2014) 7},
\href{http://arxiv.org/abs/1401.4173}{{\ttfamily arXiv:1401.4173 [hep-th]}}.

\bibitem{deRham:2010kj}
C.~de~Rham, G.~Gabadadze, and A.~J. Tolley, ``{Resummation of Massive
  Gravity},'' \href{http://dx.doi.org/10.1103/PhysRevLett.106.231101}{{\em
  Phys.Rev.Lett.} {\bfseries 106} (2011) 231101},
\href{http://arxiv.org/abs/1011.1232}{{\ttfamily arXiv:1011.1232 [hep-th]}}.

\bibitem{Hassan:2011hr}
S.~Hassan and R.~A. Rosen, ``{Resolving the Ghost Problem in non-Linear Massive
  Gravity},'' \href{http://dx.doi.org/10.1103/PhysRevLett.108.041101}{{\em
  Phys.Rev.Lett.} {\bfseries 108} (2012) 041101},
\href{http://arxiv.org/abs/1106.3344}{{\ttfamily arXiv:1106.3344 [hep-th]}}.

\bibitem{Boulware:1973my}
D.~Boulware and S.~Deser, ``{Can gravitation have a finite range?},''
\href{http://dx.doi.org/10.1103/PhysRevD.6.3368}{{\em Phys.Rev.} {\bfseries D6}
  (1972) 3368--3382}.

\bibitem{deRham:2011ca}
C.~de~Rham, G.~Gabadadze, D.~Pirtskhalava, A.~J. Tolley, and I.~Yavin,
  ``{Nonlinear Dynamics of 3D Massive Gravity},''
  \href{http://dx.doi.org/10.1007/JHEP06(2011)028}{{\em JHEP} {\bfseries 1106}
  (2011) 028},
\href{http://arxiv.org/abs/1103.1351}{{\ttfamily arXiv:1103.1351 [hep-th]}}.

\bibitem{Paulos:2012xe}
M.~F. Paulos and A.~J. Tolley, ``{Massive Gravity Theories and limits of
  Ghost-free Bigravity models},''
  \href{http://dx.doi.org/10.1007/JHEP09(2012)002}{{\em JHEP} {\bfseries 1209}
  (2012) 002},
\href{http://arxiv.org/abs/1203.4268}{{\ttfamily arXiv:1203.4268 [hep-th]}}.

\bibitem{Hassan:2011zd}
S.~Hassan and R.~A. Rosen, ``{Bimetric Gravity from Ghost-free Massive
  Gravity},'' \href{http://dx.doi.org/10.1007/JHEP02(2012)126}{{\em JHEP}
  {\bfseries 1202} (2012) 126},
\href{http://arxiv.org/abs/1109.3515}{{\ttfamily arXiv:1109.3515 [hep-th]}}.

\bibitem{Hinterbichler:2012cn}
K.~Hinterbichler and R.~A. Rosen, ``{Interacting Spin-2 Fields},''
  \href{http://dx.doi.org/10.1007/JHEP07(2012)047}{{\em JHEP} {\bfseries 07}
  (2012) 047},
\href{http://arxiv.org/abs/1203.5783}{{\ttfamily arXiv:1203.5783 [hep-th]}}.

\bibitem{Schwartz:2003vj}
M.~D. Schwartz, ``{Constructing gravitational dimensions},''
  \href{http://dx.doi.org/10.1103/PhysRevD.68.024029}{{\em Phys.Rev.}
  {\bfseries D68} (2003) 024029},
\href{http://arxiv.org/abs/hep-th/0303114}{{\ttfamily arXiv:hep-th/0303114
  [hep-th]}}.

\bibitem{Deser:2009hb}
S.~Deser, ``{Ghost-free, finite, fourth order D=3 (alas) gravity},''
  \href{http://dx.doi.org/10.1103/PhysRevLett.103.101302}{{\em Phys.Rev.Lett.}
  {\bfseries 103} (2009) 101302},
\href{http://arxiv.org/abs/0904.4473}{{\ttfamily arXiv:0904.4473 [hep-th]}}.

\bibitem{Chen:2013aha}
T.-j. Chen and E.~A. Lim, ``{Stabilization of Linear Higher Derivative Gravity
  with Constraints},''
  \href{http://dx.doi.org/10.1088/1475-7516/2014/05/010}{{\em JCAP} {\bfseries
  1405} (2014) 010},
\href{http://arxiv.org/abs/1311.3189}{{\ttfamily arXiv:1311.3189 [hep-th]}}.

\bibitem{Hawking:2001yt}
S.~W. Hawking and T.~Hertog, ``{Living with ghosts},''
  \href{http://dx.doi.org/10.1103/PhysRevD.65.103515}{{\em Phys. Rev.}
  {\bfseries D65} (2002) 103515},
\href{http://arxiv.org/abs/hep-th/0107088}{{\ttfamily arXiv:hep-th/0107088
  [hep-th]}}.

\bibitem{Mannheim:2011ds}
P.~D. Mannheim, ``{Making the Case for Conformal Gravity},''
  \href{http://dx.doi.org/10.1007/s10701-011-9608-6}{{\em Found. Phys.}
  {\bfseries 42} (2012) 388--420},
\href{http://arxiv.org/abs/1101.2186}{{\ttfamily arXiv:1101.2186 [hep-th]}}.

\bibitem{Modesto:2011kw}
L.~Modesto, ``{Super-renormalizable Quantum Gravity},''
  \href{http://dx.doi.org/10.1103/PhysRevD.86.044005}{{\em Phys. Rev.}
  {\bfseries D86} (2012) 044005},
\href{http://arxiv.org/abs/1107.2403}{{\ttfamily arXiv:1107.2403 [hep-th]}}.

\bibitem{Biswas:2011ar}
T.~Biswas, E.~Gerwick, T.~Koivisto, and A.~Mazumdar, ``{Towards singularity and
  ghost free theories of gravity},''
  \href{http://dx.doi.org/10.1103/PhysRevLett.108.031101}{{\em Phys. Rev.
  Lett.} {\bfseries 108} (2012) 031101},
\href{http://arxiv.org/abs/1110.5249}{{\ttfamily arXiv:1110.5249 [gr-qc]}}.

\bibitem{Conroy:2014eja}
A.~Conroy, T.~Koivisto, A.~Mazumdar, and A.~Teimouri, ``{Generalized quadratic
  curvature, non-local infrared modifications of gravity and Newtonian
  potentials},'' \href{http://dx.doi.org/10.1088/0264-9381/32/1/015024}{{\em
  Class. Quant. Grav.} {\bfseries 32} no.~1, (2015) 015024},
\href{http://arxiv.org/abs/1406.4998}{{\ttfamily arXiv:1406.4998 [hep-th]}}.

\bibitem{Holdom:2015kbf}
B.~Holdom and J.~Ren, ``{A QCD analogy for quantum gravity},''
\href{http://arxiv.org/abs/1512.05305}{{\ttfamily arXiv:1512.05305 [hep-th]}}.

\bibitem{Cline:2003gs}
J.~M. Cline, S.~Jeon, and G.~D. Moore, ``{The Phantom menaced: Constraints on
  low-energy effective ghosts},''
  \href{http://dx.doi.org/10.1103/PhysRevD.70.043543}{{\em Phys. Rev.}
  {\bfseries D70} (2004) 043543},
\href{http://arxiv.org/abs/hep-ph/0311312}{{\ttfamily arXiv:hep-ph/0311312
  [hep-ph]}}.

\bibitem{Hinterbichler:2015pqa}
K.~Hinterbichler and A.~Joyce, ``{A Hidden Symmetry of the Galileon},''
  \href{http://dx.doi.org/10.1103/PhysRevD.92.023503}{{\em Phys.Rev.}
  {\bfseries D92} no.~2, (2015) 023503},
\href{http://arxiv.org/abs/1501.07600}{{\ttfamily arXiv:1501.07600 [hep-th]}}.

\bibitem{Luty:2003vm}
M.~A. Luty, M.~Porrati, and R.~Rattazzi, ``{Strong interactions and stability
  in the DGP model},''
  \href{http://dx.doi.org/10.1088/1126-6708/2003/09/029}{{\em JHEP} {\bfseries
  0309} (2003) 029},
\href{http://arxiv.org/abs/hep-th/0303116}{{\ttfamily arXiv:hep-th/0303116
  [hep-th]}}.

\bibitem{Nicolis:2008in}
A.~Nicolis, R.~Rattazzi, and E.~Trincherini, ``{The Galileon as a local
  modification of gravity},''
  \href{http://dx.doi.org/10.1103/PhysRevD.79.064036}{{\em Phys.Rev.}
  {\bfseries D79} (2009) 064036},
\href{http://arxiv.org/abs/0811.2197}{{\ttfamily arXiv:0811.2197 [hep-th]}}.

\bibitem{Bergshoeff:2009hq}
E.~A. Bergshoeff, O.~Hohm, and P.~K. Townsend, ``{Massive Gravity in Three
  Dimensions},'' \href{http://dx.doi.org/10.1103/PhysRevLett.102.201301}{{\em
  Phys.Rev.Lett.} {\bfseries 102} (2009) 201301},
\href{http://arxiv.org/abs/0901.1766}{{\ttfamily arXiv:0901.1766 [hep-th]}}.

\bibitem{Muneyuki:2012ur}
K.~Muneyuki and N.~Ohta, ``{Unitarity versus Renormalizability of Higher
  Derivative Gravity in 3D},''
  \href{http://dx.doi.org/10.1103/PhysRevD.85.101501}{{\em Phys. Rev.}
  {\bfseries D85} (2012) 101501},
\href{http://arxiv.org/abs/1201.2058}{{\ttfamily arXiv:1201.2058 [hep-th]}}.

\bibitem{Muneyuki:2013aba}
K.~Muneyuki and N.~Ohta, ``{Renormalization of Higher Derivative Quantum
  Gravity Coupled to a Scalar with Shift Symmetry},''
  \href{http://dx.doi.org/10.1016/j.physletb.2013.07.054}{{\em Phys.Lett.}
  {\bfseries B725} (2013) 495--499},
\href{http://arxiv.org/abs/1306.6701}{{\ttfamily arXiv:1306.6701 [hep-th]}}.

\end{thebibliography}\endgroup

\end{document}